\RequirePackage{fix-cm}

\documentclass[smallextended]{svjour3}       
\smartqed  
\usepackage{graphicx}

\usepackage[numbers]{natbib}
\usepackage[english]{babel}
\usepackage[utf8x]{inputenc}
\usepackage[T1]{fontenc}
\usepackage{multirow}
\usepackage{enumitem}
\usepackage{fixme}
\usepackage{amssymb}
\usepackage{algorithm}
\usepackage{algorithmicx} %
\usepackage{algcompatible}
\fxsetup{
    status=draft,
    author=,
    layout=inline, 
    theme=color
}
\definecolor{fxnote}{rgb}{0.8000,0.0000,0.0000}

\usepackage{booktabs}
\usepackage{subcaption}
\usepackage{amsmath}
\usepackage{graphicx}

\usepackage[colorinlistoftodos]{todonotes}
\usepackage{hyperref}
\usepackage{pgfplots}
\pgfplotsset{compat=1.15}

\usepackage{array}
\newcommand\MyBox[2]{
  \fbox{\lower0.75cm
    \vbox to 1.7cm{\vfil
      \hbox to 1.7cm{\hfil\parbox{1.4cm}{#1\\#2}\hfil}
      \vfil}
  }
}
\begin{document}

\title{A secondary immune response based on co-evolutive populations of agents for anomaly detection and characterization
}

\titlerunning{The Reactive Artificial Bioindicator System}        

\author{Pedro Pinacho-Davidson, Matías Lermanda Sandoval, Ricardo Contreras Arriagada \and María A. Pinninghoff Junemann
}


\authorrunning{P.Pinacho-Davidson, M.Lermanda, R.Contreras \and M.A.Pinninghoff} 

\institute{Pedro Pinacho-Davidson \at
              \email{ppinacho@udec.cl}
           \and
           Matías Lermanda Sandoval \at
              \email{matlermanda@udec.cl}
          \and
           Ricardo Contreras Arriagada \at
              \email{rcontrer@udec.cl}
          \and
           María A. Pinninghoff Junemann \at
              \email{mpinning@udec.cl}
}


\maketitle

\begin{abstract}

The detection of anomalies in unknown environments is a problem that has been approached from different perspectives with variable results.  Ariticial Immune Systems (AIS) present particularly advantageous characteristics for the detection of such anomalies.  This research is based on an existing detector model, named Artificial Bioindicators System (ABS) which identifies and solves its main weaknesses.  An ABS-based anomaly classifier model is presented, incorporating elements of the immune system AIS.  In this way, a new model (R-ABS) is developed which includes the advantageous capabilities of an ABS plus the reactive capabilities of an AIS to overcome its weaknesses and disadvantages. The RABS model was tested using the well-known DARPA'98 dataset, plus a dataset built to carry out a greater number of experiments. The performance of the RABS model was compared to the performance of the ABS model based on classical sensitivity and specificity metrics, plus a response time metric to illustrate the rapid response of R-ABS relative to ABS. The results showed a better performance of R-ABS, especially in terms of detection time.

\end{abstract}

\section{Introduction}

There are many industrial application in which the detection of anomalies and their corresponding characterization is vital to achieve operational continuity. This approach is advantageous when the failure conditions are not known in advance. In these cases, the classification system cannot be trained before entering the production stage. Some examples of these scenarios are: machine fail detection over unknown operational environments \cite{kuang2018stable}, new object recognition over hyper-spectral imagery \cite{REF5}, and the treatment of unknown cybersecurity threats \cite{garcia2009anomaly, ahn2014big, narayanan2018early}. 

Different machine learning approaches have been applied to the problem of detection and characterization of anomalies.  Heuristic approaches to infer the type of anomalies \cite{REF1}, flow-based statistical analysis \cite{REF2}, wavelet transformation based \cite{REF3}, cluster approaches \cite{REF4}, generative adversarial networks \cite{REF6}, and different proposals grouped in the family of Artificial Immune Systems based algorithms (AIS) \cite{fernandes2017applications}. 

AIS presents  inherent  capabilities for the treatment of the anomaly detection and characterization problem  implementing diverse strategies and algorithms based on the Biological Immune System of the mammals (BIS) \cite{gorriz2020artificial}. One of the most popular AIS algorithm is the Negative Selection Algorithm (NSA), proposed by Forrest (\citet{1}), which is based on the immune cells selection process developed in the thymus.  

Another important algorithm is the Clonal Selection (CSA), proposed by Burnet (\citet{22}). The most known implementation of CSA is CLONALG (\citet{55}), it is a pattern recognition proposal, based on mutation for diversification of the best detectors, and the removal of the less representatives. A fundamental, and relatively new algorithm, in the field of immunology is the Danger Theory Algorithm (DT), proposed by Matzinger (\citet{20,66}). DT removes the need to model the self-behavior --normal behavior--, but it requires the definition of environmental signals associated with the anomalies. 

In this work, we consider a different perspective on immunity.   We understand the system to be protected as a dynamic one, whose state is structurally adjusted, continuously and plastically, during its operation adding new elements and removing the old ones. The addition and removal of the elements is controlled by its internal dynamics of the system. 

This behavior is associated with the concept of \textbf{endogenous double plasticity} (\citet{EDP}) and may be achieved through simple heuristics such as the compensation of weak elements, the preservation of diversity, and the elimination of the redundancy, i.e.,  the establishment of a balance using ecological mechanism.  

Immunity is understood as the dynamic interaction of a population of agents.  In this context we deal with the modeling task, using the \textbf{Agent-Based Modeling} (ABM) approach. A previous work entitled \textbf{Artificial Bioindicators System} (ABS) (\citet{AB-EIA}) shows the evidence of the usefulness of the population of agents for the detection of anomalies in the context of computer security.

Our proposal is centered on achieving other advantage associated with BIS, by exploiting the \textbf{characterization-memorization} of threats and the possibility of a quick \textbf{secondary immune response}. For this purpose, we incorporate the DT concept, allowing the control of the relationship between the original agent's population and a new kind of reactive agents. These new agents are deployed for the implementation of the secondary immune response of the system.\\

The objective of this article is to report the progress of the development of the \textbf{Reactive Artificial Bioindicators System} (R-ABS), with a significant improvement of the capacities of the ABS (\citet{AB-EIA}). The original model proposes an IDS inspired by an ecological approach to BIS, evolving a population of agents learning to survive in their environment. Thus the adaptation process allows the transformation of the agent population into bioindicators which are capable of detecting system anomalies. We use this work as a base, and we propose an extension for achieving the characterization of the threats and improving the overall response time of the system. \\  

In this work, we report four contributions; the first one is the simplification the original ABS model. The second contribution is the extension of the model for achieving threat characterization skills over the system. The third one provides a mechanism for protecting the self-knowledge of the system, avoiding profile contamination. The fourth improvement is the overall reduction of the time for detecting threats to which the system was previously exposed. \\ 

The paper is structured as follows.  After this introduction,  the second section presents the ABS model and its characteristics. Then, section three describes the R-ABS proposal, and section four describes the experiments developed followed by the results and the conclusion in the last two sections of the article.

\section{The ABS Model}

The ABS (\citet{AB-EIA}) was built under an ecological point of view, from the BIS and tested in a cybersecurity scenario. In this context, ABS deals with the problem of \textbf{network intrusion detection}.  Consequently, the system agents adapt to a normal behavior of the network data.  This allows the detection of an attack in the same way the natural bioindicators do, because these agents are not adapted to the hostile environment. ABS uses a population of agents very sensitive to detection anomalies. For this reason, these agents are called \textbf{Artificial Bioindicators (AB)}. 

The system is represented as a 2-dimensional world with two entities: {\em agents} and {\em particles}. Agents act like bioindicators. Changes in the environment disturb these agents. The environment corresponds to the traffic over the network data.  The particles represent the characteristics of the real network traffic, on the two-dimensional world. 

Agents are static entities initially created at random;  these agents undergo an evolutionary processes that allow them to adapt and ensure that only the most suitable survive. Each agent has two key parameters associated with it: 

\begin{itemize}
\item \textbf{Energy} ($E_{x}$): Indicates the amount of energy that the agent possesses, this energy is used to keep the agent's metabolism activated and for reproduction purposes.

\item  \textbf{Genetic code} ($\theta$): It is a vector that determines the agent's qualities, in other words,  the agent's ability to survive some characteristic of the environment.  Each $i \in \theta$  represents a single gene, and each gene represents one of the characteristics observed in the network traffic.
\end{itemize} 

Agents require a certain level of energy  ($E_{r}$) to reproduce, which is a parameter.  Agents reproduce asexually, activating a mutation operator on their genetic code, which consists of a random exchange of two elements in their genome, this small variation determines the affinity of the child with the environment.\\ 

\textbf{Particles} are created by the system when a specific characteristic is found in network packets.
The system adds the particles on one side of the world (according to the definition of its topology).  The particles are then moved, one step at a time, to the other side, until there are no more particles left on the first chosen side.
Particles have two key characteristics:  {\em type} and {\em capacity}. The type describes the characteristic of the traffic represented by the particle, and the capacity represents the number of agents capable of absorbing the particles with which they collide.\\ 

When a particle $p$ collides with an agent, its energy level can be modified according to the following equation: 

\begin{equation}
N_{x}=\sum_{p}(\Phi - \epsilon* i_p)
\end{equation}

$N_{x}$ is the amount of energy that a particle delivers to an agent $x$,  $\Phi$ is the maximum energy value that a particle can deliver to an agent, $i_p$ is an index which represents the type of particle that can be found in the genetic code of the agent $x$, and $\epsilon$ represents a penalizing factor for the energy value.  The collision of a particle with an agent increases or decreases the energy of the agent, depending on the genetic affinity between the agent and the particle. \\ 

As an example, consider a system with five different particles.  Each agent has a genetic vector, which contains a permutation of the five particles.  In this way, if an agent $A$ has a vector $A_v=(e,a,c,b,d)$ and the arrival of a package $P$ to the network introduces the particles $e$ and $b$, the nutritional value associated to $P$  given $A$ will be: $N_A=(\Phi - \epsilon * i_e) + (\Phi - \epsilon* i_b)$, where $i_e$ is  $0$ and $i_b$ is 3. In this scenario, the particle $e$ produces the maximum nutritional profit for A. On the other hand, the particle $b$ is less beneficial and can even be toxic (negative value of $N$) depending on the value of $\epsilon$.\\

The ABS model presents the following important issues: 
\begin{enumerate}

\item \textbf{Lack of reactive capabilities}: The original ABS model does not have built-in detection capabilities; detection is done through the use of an external observation process. 

This process receives the agent's cardinality and their energy level as input.  Then, this tool uses this information for classification purposes. The output can be one of these answers: {\em normal situation} or {\em system under attack}. Therefore, the attack report is not triggered by the agent's reaction.\\

\item \textbf{The problem of the retraining of agents:}    When an attack persists over time, agents tend to adapt to it. This is an undesirable side effect since the agents can become contaminated with the attack. Its function should be to characterize a normal operation of the system.

\item \textbf{Slow reaction capability:}  The dynamics of the system is given by a set of adaptive agents, and the classification process is based on the disturbance that the network attack causes in the agents.  Since the disturbance is analyzed by an external tool, that considers trend patterns that involve some parameters of the system, the final reaction can be slow.

\item \textbf{ Lack of learning capacities} The original model does not provide a  learning mechanism useful for the treatment of repeated attacks.
\end{enumerate}

\section{The R-ABS Model}

This section describes a proposal for dealing with the four problems identified in the ABS model. To solve these problems, the new model must satisfy the following requirements:

\begin{itemize}

\item All reactions are triggered by the agents of the model. No external tools participate in this action, nor do global metric values of the model.  In this sense, the new proposal does not alter the implicit rules on the existing relations between neighbors; and that correspond to the functionalities activated by the emergency.  This is how the problem of {\em lack of reactive capacities} of the ABS model is solved.

\item The new model tackles the {\em problem of the retraining of agents}, providing a mechanism for the protection of the self-characterization, achieved by the implementation of new memory entities and constraints relationships, to encourage or inhibit their creation.


\item The new proposal as an AIS has two levels of response to threats. First, an innate and generic response for unknown attacks and second, an adaptive response for known attacks. In this second level,  the characterization of the unknown threats achieves the speed-up of the reactions. Allowing the detection by signatures and not only by anomaly detection. In this way,  R-ABS tries the {\em slow reaction and the lack of learning capacities} of ABS.

\end{itemize}

\begin{figure}
    \centering
    \includegraphics[width=0.9\textwidth]{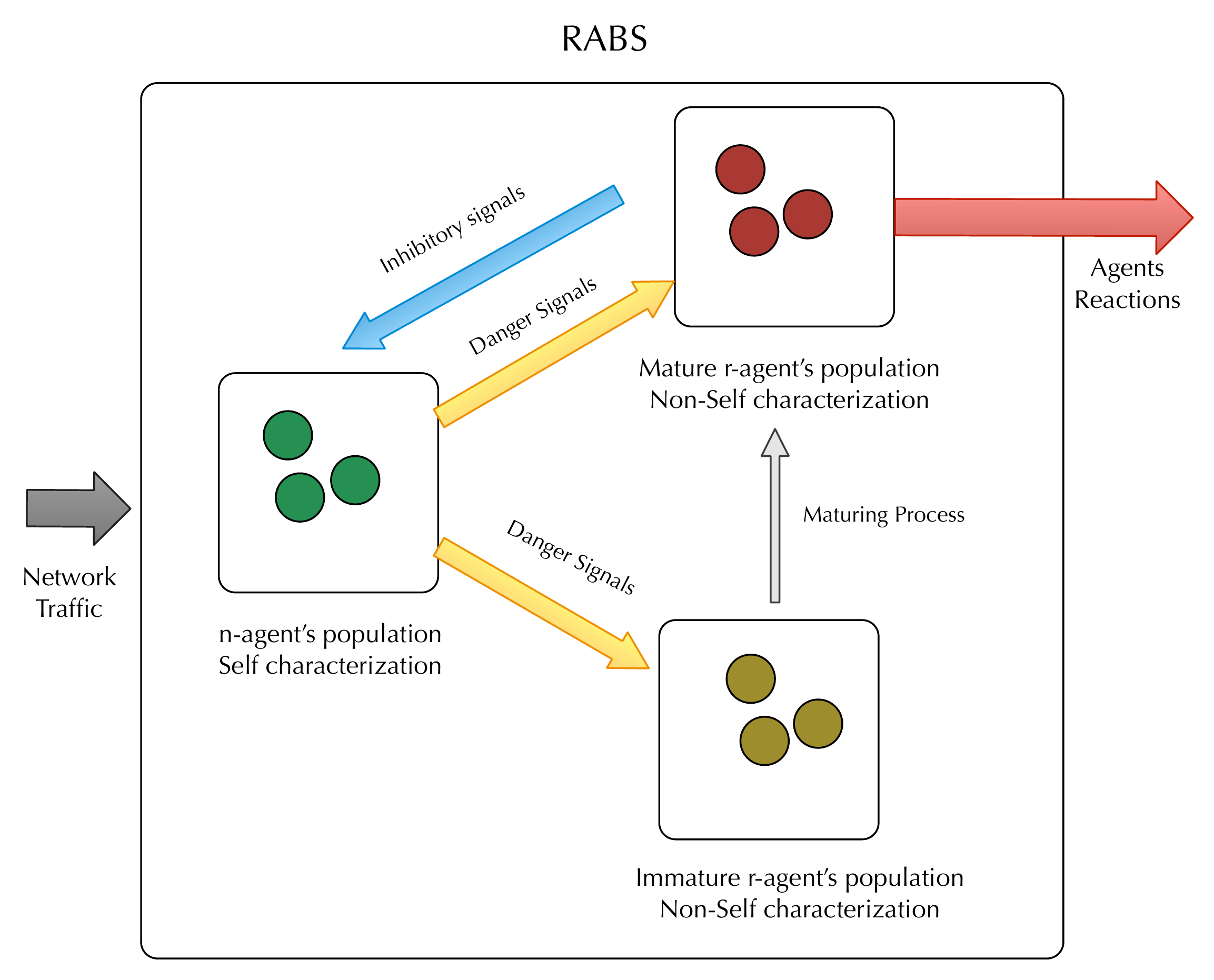}
    \caption{Main Model of R-ABS}
    \label{fig:RABS}
\end{figure}

In Figure \ref{fig:RABS}, it is shown the R-ABS response on the same ABS action scenario.   The structure is composed by a set of different agents populations, and the network traffic is captured by a sniffer that is in charge of feeding the system.

R-ABS has two types of evolutionary agents.  The first type is quite similar to the original definition of agents in the ABS model. These agents are called n-agents (agents of normality) in the R-ABS model.  The n-agents are defined to characterize the normal behavior of the system being monitored (Self) and each R-ABS model maintains a population of these agents.  The second type of agent corresponds to the r-agents (reactive agents); this type of agent is designed to learn the threat's behavior. 
R-ABS can deploy and maintain several r-agents populations, one population for each different threat occurrence.

There are different signals that communicate different agent's populations.  in this way, perturbations over n-agents produce danger signals, and these signals are used for the creation and consolidation of populations of r-agents. It is important to highlight that the creation of multiples r-agents populations does not imply a computing or memory overhead, because only one population is created per event, and when a r-agents population is already trained, it maintains only a few consolidated agents for a straightforward pattern recognition procedure. \\

The presence of an anomaly (perturbation of n-agents) triggers the activities of r-agents in the same way that innate immunity activates adaptive immunity. Mature (Consolidated) r-agents can produce other types of signals when known attacks activate them. These signals have the purpose of inhibiting the evolution of n-agents, preventing the contamination of normal (Self) behavior characterization. In general terms, the relationship between signals and agents populations can be described by means of the following rules:
\\

\scalebox{1}{
(a) $\blacktriangledown$(n-agents)$\wedge\blacktriangle$(r-agents)$\overset{produce}{\rightarrow}$ \{full reaction\}  
}
\\
\scalebox{1}{
(b) $\blacktriangledown$(n-agents)$\wedge\blacktriangledown$(r-agents)$\overset{produce}{\rightarrow}$ \{new population, soft reaction\} 
}
\\
\scalebox{1}{
(c) $\blacktriangle$(n-agents)$\wedge\blacktriangle$(r-agents)$\overset{produce}{\rightarrow}$ \{full reaction, i-NSA  \}}
\\
\scalebox{1}{
(d) $\blacktriangle$(n-agents)$\wedge\blacktriangledown$(r-agents)$\overset{produce}{\rightarrow} \emptyset$  }
\\

This set of four rules explains the primary behavior of R-ABS. The rules are not triggered by an external tool but by the direct action of the agents. In rule (a), R-ABS is operational, and under attack. In this situation, the n-agents lost their affinity to the incoming network traffic. Also, one of the r-agents population presents high-affinity levels with the incoming traffic, producing the reaction of the r-agents over the recognized threat. Rule (b) is triggered when the R-ABS is exposed to a new threat.  In this way, the n-agent affinity to the traffic decreases, but none of the r-agents population has high-affinity with the incoming attack. In this situation, R-ABS deploys a new r-agent population and adapts it to the incoming threat.  In this new-threat scenery, R-ABS can produce a soft reaction in the same way that innate immunity produces a nonspecific reaction in BIS. Rule (c) shows an inconsistency between the R-ABS population. While the r-agents affinity levels mark the presence of a threat, the n-agents population is not affected by the same perturbation. In this situation, R-ABS apply a non-scheduled variation of the Negative Selection Algorithm (i-NSA) to solve the inconsistency, and perform the full reaction of the r-agents. i-NSA is described in the next subsection. Finally, the rule (d) denotes the state of R-ABS in the absence of attacks. In this condition, the affinity of the n-agents is high and does not exist a population of r-agent with high affinity over the incoming traffic.

\subsection{Inverted Negative Selection Algorithm}

The Negative Selection Algorithm NSA was introduced by Forrest in \cite{1}. The NSA's primary goal is to allow the safe creation of new antigens detectors, destroying detectors that classify the self-behavior as a threat before they become mature. This algorithm is based on the immutability of the Self. We consider a different premise in R-ABS. The normal behavior of the systems is mutable, and it changes over time, but a specific threat is static. In this way, the exposure to a variation of a known attack triggers the generation of new r-agents populations to characterize this particular unknown hostile behavior. In other words, threat characterizations are, by definition, immutable. \\

R-ABS takes as reference the known threat for the destruction of harmful n-agents, this is, with high-affinity to the characterization of attacks provided by r-agents. In this way, inverted Negative Selection (i-NSA) is similar to NSA but cleaning the Self repertory taking as reference the non-Self definitions, and not cleaning the threat detectors repertory based on the Self repertory as the original NSA algorithm. 

\subsection{Model Simplification}

We conduct a simplification of the original model, from a $2D$ world to a $1D$ topology. The main reason for doing this is that the ABS model does not justify the two dimensions need. In this sense,  ABS does not take advantage of the 2D-world properties like distance between objects, movements, and agents' emergent clusters. We suppose a reduction of the computational effort by re-thinking the ABS idea over a 1D Topology, with a fixed number of agents, fewer parameters, and more simple interaction rules between the model elements. Besides that, ABS has other not informed settings values such as the size of the world $(range-in-X, range-in-Y)$, the speed of the particles in the model, and maybe other virtual world parameters that can affect the model behavior. \\

\subsection{R-ABS over Cybersecurity Domain}

Technically, R-ABS is a {\em Network-based Intrusion Detection System} (NIDS), a system designed to detect network threats. In this way, The required input data is taken from a monitoring system generating an output report, useful for stop or mitigate the incoming hostile behavior. R-ABS was conceived as a symbiotic structure coupled with its host. In this way, R-ABS tries to maintain the network homeostasis for surviving and preserve the host's health. 
The input used by R-ABS is the network traffic of the monitored organizational network. We consider a set of binary characteristics extracted from headers of the Transport and Network layer of (TCP-UDP)/IP connections. We also consider SANS  reference over the most attacked service ports for individualizing some critical services \cite{AB-EIA}. The selection of these characteristics is justified because they are used by ABS and allow a fair comparison between the algorithms.  The detail for input information used by R-ABS is the following:

\begin{itemize}
    \item \textbf{Reserved bits and network flags:} RB, MF, DF, F1, F2, URG, ACK, PSH, RST SYN, FIN.
    \item \textbf{Most attacked port (SANS):} Telnet, SSH, FTP, Netbios, Rlogin, RPC, NFS, NNTP, Lockd, Netbios, Xwin, DNS, LDAP, SMTP, POP, NTP, IMAP, HTTP, SSL, px, Serv, Time, TFTP, Finger, lpd, Syslog, SNMP, bgp, Socks.

\end{itemize}

Finally, we create a single binary vector $ V $ of length $ lv $ for each incoming network packet, where $ lv $ is the number of different traffic characteristics. Each position in the vector is associated with a specific traffic property.\\

\subsection{The agents of the R-ABS Model}

R-ABS preserves the idea of agents that act bioindicators.  Thus, the environment (network traffic) establishes a pressure mechanism that causes the population of agents continually adapts to this changing environment. In this context, the fittest individuals are those with higher levels of energy. All agents implemented in R-ABS have evolutive properties. For this purpose, each agent holds a genetic vector. This vector $ G $ establishes the affinity of the agents with the environment. The genetic representation of $ G $ is the same as the binary vector $V$ for the characterizing the incoming network traffic.

The evolving mechanism of R-ABS is described through Algorithms 1  and \ref{algo:evolution-r}. The first one represents the behavior of the n-agents population, meanwhile the second algorithm represents the r-agents operation. Both procedures are very similar between them, and they are codependent. In this way, the n-agents population activates the evolution of the set of r-agents, and the r-agents populations can inhibit the n-agents population's change. These algorithms are activated whenever a new network package is captured.

 In detail, Algorithm 1 shows that the n-agents population $A$ is exposed to each network packet ($p_{cap}$), then, in line 4, the number of agents with energy below $energy_{thres}$ is calculated and stored in DS (danger signals), and the entire r-agent population inspects it because DS trigger their execution. In line 5, the algorithm checks if the Attack level evidence ($AS$) is under a threshold $AS_{thres}$. When the condition is true, the population continues evolving normally; in the other case, the algorithm inhibits the evolving process for preventing the contamination of the n-agents population with attack behavior. When the protection condition is satisfied, the algorithm selects a parent's group ($A'$). The chosen agents, from $A$, must have an energy level above a threshold ($energy_{thres} $). This is, agents already adapted to the normal network conditions. 
 
 
 In the same way, the algorithm selects the a second group of agents ($A*$) from $A$, using the threshold ($Mem_{thres}$) in an elitist form ($Mem_{thres} >> energy_{thres}$). The Agents in $A*$ are considered matured because of their high affinity with the normal incoming traffic. Algorithm 1 defines $r$ as the number of new agents to be created and stored in $A_n$; the value of $r$ is determined using a proportion ($R$) of the total population of n-agents and considering the constraint of at least replacing dead agents. In lines 11 and 12, two parents ($p_1, p_2$) are selected from $A'$ using {\em Fitness Proportionate Selection} ($FPS$) and combined in ($o1, o2$) using a uniform crossover ($X_{uni}$) \cite{Michalewicz2004} , then the new offsprings are added to $A_n$. This process is repeated until obtaining a number of new agents equal to, or larger than, $r$. In line 15, the algorithm step represents the replacement mechanism for the n-agents. The policy is straightforward, and it consists of changing the weakest n-agents in terms of their energy level with the new agents in $A_n$. The normal behavior of the system can change over time. In this way, R-ABS must provide a mechanism to update the representation of the Self. This process is achieved through the function update (line 16) that replaces the weakest mature r-agents (memory cells)  in $A^{**} $ with new high energy elements in $A^{*}$.

The evolving mechanism of r-agents is presented in Algorithm 2. The algorithm is quite similar to the one previously presented. The main difference is that the evolution and the evaluation of the r-agents are allowed only in the presence of danger signals ($DS$). In this way, the algorithm takes as input the value of $DS$ determined by the n-agents population. When this value achieves the threshold level ($DS_{thres}$), r-agents are exposed to the network traffic ($p_{cap}$) and evolve with the same rules that govern n-agents evolution. Another significative difference is that r-agents does not have an update mechanism for mature agents. Thus, when one population of r-agents characterizes an attack scenario, it stops evolving and it is only used to establish an already known attack.  The r-agents are responsible for computing the value of attack signals ($AS$) used by the n-agents population to inhibit the evolving process in case of attacks (line 5).\\

\begin{algorithm}[t]
\caption{Evolving mechanisms of n-agents in R-ABS} \label{evolutionn}
\begin{algorithmic}[1]
\STATE {\bf input:} $A,energy_{thres},R,p_{cap},AS$ 
\STATE {\bf input:} $AS_{thres}, Mem_{thres}$
\STATE $expose(A, p_{cap})$
\STATE $DS =|\left \{ a \in A^{**}  | a_{energy} < energy_{thres}  \right \}|$
\IF{ $AS < AS_{thres}$}
\STATE $A':=\left \{ a \in A  | a_{energy} \ge energy_{thres}  \right \}$
\STATE $A^{*}:= \left \{ a \in A  | a_{energy} \ge Mem_{thres} \right \}$
\STATE $A_n:=\emptyset$
\STATE $r:=argMax(deadAgents,R\cdot|A|)$
\WHILE{$|A_n|<r$}
\STATE $p_1,p_2:=FPS(A')$
\STATE $o_1,o_2:=X_{uni}(p_1,p_2)$
\STATE  $A_n:=A_n\cup \{ o_1, o_2\}$
\ENDWHILE
\STATE $replace(A,An, policy(lower\_energy))$
\STATE $update(A^{**},A^{*}, policy(lower\_energy))$
\ENDIF
\end{algorithmic}
\end{algorithm}

\begin{algorithm}[t]
\caption{Evolving mechanisms of r-agents in R-ABS} \label{algo:evolution-r}
\begin{algorithmic}[1]
\STATE {\bf input:} $A,energy_{thres},R,p_{cap},DS$ 
\STATE {\bf input:} $DS_{thres}, Mem_{thres},AS_{thres}$
\IF{ $DS > DS_{thres}$}
\STATE $expose(A, p_{cap})$
\STATE $AS =|\left \{ a \in A^*  | a_{energy} \ge AS_{thres}  \right \}|$
\STATE $A':=\left \{ a \in A  | a_{energy} \ge energy_{thres}  \right \}$
\STATE $A^{*}:=A^{*} \cup \left \{ a \in A  | a_{energy} \ge Mem_{thres}  \right \}$
\STATE $A_n:=\emptyset$
\STATE $r:=argMax(deadAgents,R\cdot|A|)$
\WHILE{$|A_n|<r$}
\STATE $p_1,p_2:=FPS(A')$
\STATE $o_1,o_2:=X_{uni}(p_1,p_2)$
\STATE  $A_n:=A_n\cup \{ o_1, o_2\}$
\ENDWHILE
\STATE $replace(A,An, policy(lower\_energy))$
\ENDIF
\end{algorithmic}
\end{algorithm}

\section{Experimental Evaluation}
The experimental setup was designed to check out the on-line attack learning capabilities of R-ABS and the possible improvement in the response speed of the model for previously seen attacks, not for validating R-ABS as a full operational IDS.  In order to comply with these requirements, we carry out two types of experiments. The first one establishes a comparison between R-ABS and ABS models, and the second one was designed for determining the R-ABS proposal's advantages. All experiences are structured using a subset of the following phases:


\begin{itemize}
\item \textbf{Phase 1}: The model is exposed to normal network traffic.  The goal is to achieve the initial adaptation to the protected network's ordinary operation condition.
\item \textbf{Phase 2}:  The model is exposed to an unknown attack to evaluate its response.


\item \textbf{Phase 3}: Network traffic is changed back to normal conditions.
\item \textbf{Phase 4}:  The model is exposed again to the attack presented in Phase 2 for testing the algorithms' secondary immune response.
\end{itemize}

To deal with the non-deterministic behavior of the algorithms under evaluation, every experiment is executed ten times. The experimental dataset used is composed of elements from DARPA'98 and some complementary normal and hostile traffic captured for this work generated with \textit{GoldenEye} \footnote{https://sourceforge.net/projects/goldeneye/}. The details about the used testbed can be inspected in the following table:\\


%
\begin{table}
\caption{Experimental Dataset Composition}
\centering
\scalebox{1.2}{
\begin{tabular}{ l|l|r }

\hline
\textbf{Type of Attack} & \textbf{Name} & \textbf{Packets} \\ \hline
\multirow{4}{*}{Denial of Service  (DoS)} & {\em land} & 1100 \\
 & {\em syslog} & 1002 \\
 & {\em udp-storm} & 1667 \\
 & {\em custom$^{*}$ } & 3000 \\ \hline
 
\multicolumn{2}{ c| }{DoS total} & 6769 \\
\hline

\multirow{2}{*}{Remote to Local (R2L)} & {\em http-tunnel } & 1196 \\
 & {\em sendmail} & 1548 \\ \hline
 
 \multicolumn{2}{ c| }{R2L total} & 2744 \\
\hline

\multirow{4}{*}{User to Root (U2R)} & {\em sql} & 1026 \\
 & {\em ffbconfig} & 1533 \\
 & {\em ps} & 703 \\
\hline

\multicolumn{2}{ c| }{U2R total} & 3262 \\
\hline

Normal traffic & normal & 8618 \\
\hline

\multicolumn{2}{ c| }{\textbf{Total}} & 21393 \\
\hline
\end{tabular}
}
\end{table}


\subsection{ ABS v/s R-ABS Baseline Experiment}

ABS model has two important weaknesses already described (the problem of the retraining of agents and lack of learning capacities). It was taken into account for establishing a fair comparison between R-ABS and ABS in terms of pure anomaly detection capacity. Thus, we produce an experiment with only the two first phases already described. Also, Phase 2 is interrupted when the ABS's agents begin to adapt to the abnormal behavior. Then, the same traffic pattern was applied over R-ABS. This baseline experiment has the same structure described in the original ABS paper \cite{AB-EIA}.

\subsection{ ABS v/s R-ABS Extended Comparison}
For a better understanding of the differences between ABS and R-ABS, two additional experiments were carried out. The first one is two phases too, but without the extension constraints of the attack phase (Phase 2). In this sense, this experiment does not protect from the agent retraining problem. The second experiment is a simple extension of the previously described. This experience adds Phase 3. Thus, this experiment verified the classifiers' response when the normality is restored.

\subsection{ R-ABS Capabilities Evaluation}
Because R-ABS has learning capabilities not available in ABS, we perform an additional experiment. This experience uses the four defined phases.  Thus, R-ABS is exposed twice to each particular attack for comparing the behavior between an unknown attack and a known one for the second exposure. 

\section{Results}

The Results of the experiments for the ABS and R-ABS models are presented using confusion matrices. Thus, the result tables show the correlation between the {\em Actual} or correct status of each evaluated network package versus the {\em Predicted} status informed by the models. This status can be Normal ($N$) and Attack ($A$). In the case of an {\em Actual} Attack package informed by the model (Predicted) as Attack, it will increment the $TP$ (True Positive counter); but if the predicted status is Normal, it will increment the $FN$ (False Negative counter). The $TN$ (True Negative) and $FP$ (False Positive) are similarly calculated when the actual status of the evaluated package is Normal. We used the previous values to compute the {\em sensitivity } $Sens=TP/(TP+FN)$ and the {\em specificity} $Spec=TN/(TN+FP)$.

Table \ref{baseline} shows the baseline experiments results with a Phase 2 that does not compromise the ABS agents' integrity. In this situation, ABS performs better than R-ABS with higher sensitivity for a similar specificity, this scenario shows the strength of ABS as an anomaly detector but does not highlight its weaknesses.

\begin{table}[!htb]
    \caption{Baseline Results}
    \label{baseline}
    \begin{subtable}{.55\linewidth}
      \centering
        \caption{ABS Baseline}
        \scalebox{1.3}{
        \begin{tabular}{cc|cc}
        \multicolumn{1}{c}{} &\multicolumn{1}{c}{} &\multicolumn{2}{c}{Predicted} \\ 
        \multicolumn{1}{c}{} & 
        \multicolumn{1}{c|}{} & 
        \multicolumn{1}{c}{$A$} & 
        \multicolumn{1}{c}{$N$} \\ \hline
        \multirow[c]{2}{*}{\rotatebox[origin=tr]{90}{Actual}}
        & $A$  & 20.9 & 9.1   \\[1.5ex]
        & $N$  & 53.4   & 756.6 \\ \hline
        & &Sens:& 0.70\\
        & &Spec:& 0.93\\
        \\

        \end{tabular}
        }
    \end{subtable}%
    \begin{subtable}{.5\linewidth}
      \centering
        \caption{R-ABS Baseline}
        \scalebox{1.3}{
        \begin{tabular}{cc|cc}
        \multicolumn{1}{c}{} &\multicolumn{1}{c}{} &\multicolumn{2}{c}{Predicted} \\ 
        \multicolumn{1}{c}{} & 
        \multicolumn{1}{c|}{} & 
        \multicolumn{1}{c}{$A$} & 
        \multicolumn{1}{c}{$N$} \\ \hline
        \multirow[c]{2}{*}{\rotatebox[origin=tr]{90}{Actual}}
        & $A$  & 18.6 & 1.4   \\[1.5ex]
        & $N$  & 58.9   & 751.5 \\ \hline
        & &Sens:&0.62\\
        & &Spec:&0.93\\
        \\
        \end{tabular}
        }
    \end{subtable} 

\end{table}

Table \ref{extended} (a),(b)  shows the results achieved when the experiments use a Phase 2 with arbitrary length (unique difference with the baseline experiment). In (a), the table shows a dramatic decrement in ABS's sensitivity from  0.7 to 0.33 considering the baseline experiment. Meanwhile, R-ABS (b) increments its sensitivity from 0.62 to 0.71 with a very controlled decrement of the specificity from 0.93 to 0.9. The results are similar for experiments that include the restoration of the normality (Phase 3) in (c) and (d).

This decrease in performance for ABS is due to the innate capabilities of the agents to adapt to anything, even an attack, rendering the model mostly indifferent to the ongoing attack, R-ABS, on the other hand, has does not have this problem.

\begin{table}[!htb]
    \caption{Extended Comparison Results}
    \label{extended}
\begin{subtable}{.5\linewidth}
      \centering
        \caption{ABS 2-phases}
        \scalebox{1.3}{
        \begin{tabular}{cc|cc}
        \multicolumn{1}{c}{} &\multicolumn{1}{c}{} &\multicolumn{2}{c}{Predicted} \\ 
        \multicolumn{1}{c}{} & 
        \multicolumn{1}{c|}{} & 
        \multicolumn{1}{c}{$A$} & 
        \multicolumn{1}{c}{$N$} \\ \hline
        \multirow[c]{2}{*}{\rotatebox[origin=tr]{90}{Actual}}
        & $A$  & 46.7 & 92.8   \\[1.5ex]
        & $N$  & 53.3   & 756.7 \\ \hline
        & &Sens:& 0.33\\
        & &Spec:& 0.93\\
        \\
        \end{tabular}
        }
        
    \end{subtable}%
    \begin{subtable}{.5\linewidth}
      \centering
        \caption{R-ABS 2-phases}
        \scalebox{1.3}{
        \begin{tabular}{cc|cc}
        \multicolumn{1}{c}{} &\multicolumn{1}{c}{} &\multicolumn{2}{c}{Predicted} \\ 
        \multicolumn{1}{c}{} & 
        \multicolumn{1}{c|}{} & 
        \multicolumn{1}{c}{$A$} & 
        \multicolumn{1}{c}{$N$} \\ \hline
        \multirow[c]{2}{*}{\rotatebox[origin=tr]{90}{Actual}}
        & $A$  & 98.9 & 30.6   \\[1.5ex]
        & $N$  & 58.5   & 751.6 \\ \hline
        & &Sens:& 0.71\\
        & &Spec:& 0.90\\
        \\
        \end{tabular}
        }
    \end{subtable} 

 \begin{subtable}{.5\linewidth}
      \centering
        \caption{ABS 3-phases}
        \scalebox{1.3}{
        \begin{tabular}{cc|cc}
        \multicolumn{1}{c}{} &\multicolumn{1}{c}{} &\multicolumn{2}{c}{Predicted} \\ 
        \multicolumn{1}{c}{} & 
        \multicolumn{1}{c|}{} & 
        \multicolumn{1}{c}{$A$} & 
        \multicolumn{1}{c}{$N$} \\ \hline
        \multirow[c]{2}{*}{\rotatebox[origin=tr]{90}{Actual}}
        & $A$  & 48.6 & 96.4   \\[1.5ex]
        & $N$  & 115.2   & 1564.8 \\ \hline
        & &Sens:& 0.34\\
        & &Spec:& 0.93\\
        \\
        \end{tabular}
        }

    \end{subtable}%
    \begin{subtable}{.5\linewidth}
      \centering
        \caption{R-ABS 3-phases}
        \scalebox{1.3}{
        \begin{tabular}{cc|cc}
        \multicolumn{1}{c}{} &\multicolumn{1}{c}{} &\multicolumn{2}{c}{Predicted} \\ 
        \multicolumn{1}{c}{} & 
        \multicolumn{1}{c|}{} & 
        \multicolumn{1}{c}{$A$} & 
        \multicolumn{1}{c}{$N$} \\ \hline
        \multirow[c]{2}{*}{\rotatebox[origin=tr]{90}{Actual}}
        & $A$  & 104.5  & 30.5   \\[1.5ex]
        & $N$  & 170.4   & 1509.6 \\ \hline
        & &Sens:&0.72\\
        & &Spec:&0.90\\
        \\
        \end{tabular}
        }
    \end{subtable}

\end{table}

Table \ref{RABS} and Table \ref{Reaction}  presents the results for the most relevant experiment over R-ABS. This experiment uses the four phases defined. In this sense, the experience allows us to known the effects of the activation mechanism using danger signals to deploy and mature r-agents populations. Also, allows us to observe the impact of the evolution inhibition process of the n-agents population induced by attack signals launched by r-agents that prevent the n-agent contamination. Additionally, the experiment evaluates the learning capabilities of R-ABS and the effect on the detection speed for the exposition to a previously detected and characterized threats.  It is interesting to note from Table \ref{RABS} that the sensitivity of R-ABS in the 4-phases experiment increase when compared to the {\em 3-phases} experiment. This information can be correlated with Table \ref{Reaction}, where R-ABS did not detect the attacks {\em ffbconfig} and {\em ps} in the first exposition (RABS column), but it was detected in the second exposure (RABS* column).

\begin{table}[!htb]
    \caption{R-ABS Evaluation Results}
     \label{RABS}
    
     \begin{subtable}{1\linewidth}
      \centering
        \caption{R-ABS 4-phases}
        \scalebox{1.3}{
        \begin{tabular}{cc|cc}
        \multicolumn{1}{c}{} &\multicolumn{1}{c}{} &\multicolumn{2}{c}{Predicted} \\ 
        \multicolumn{1}{c}{} & 
        \multicolumn{1}{c|}{} & 
        \multicolumn{1}{c}{$A$} & 
        \multicolumn{1}{c}{$N$} \\ \hline
        \multirow[c]{2}{*}{\rotatebox[origin=tr]{90}{Actual}}
        & $A$  & 224.2 & 45.8   \\[1.5ex]
        & $N$  & 201   & 1479 \\ \hline
        & &Sens:&0.80\\
        & &Spec:&0.88\\
        \\
        \end{tabular}
        }
    \end{subtable}

\end{table}

Table \ref{Reaction} shows the reaction times of ABS and R-ABS for different attacks from the dataset. The results are shown in terms of the packages needed for raising the alert of attack. In this table, the label \textbf{RABS} indicates the first exposure to the specific threat, and \textbf{RABS}$^*$ for the second exposure to the same attack. Anomaly detection is faster in ABS than R-ABS. Nevertheless, R-ABS performs two different operations during the first exposure: anomaly detection and anomaly characterization. The learning capability of R-ABS allows a quasi-instantaneous secondary response for a previously characterized threat, where the classifier requires only two network packages for the attack identification in the average case.

\begin{table}
\caption{Comparison of the Reaction Time}
\centering
\label{Reaction}
\scalebox{1.2}{
    \begin{tabular}{ l|c|c|c }
    \hline
    \textbf{Attack Name} & \textbf{ABS} & \textbf{RABS} & \textbf{RABS}$^*$ \\ \hline
    land & 80.5 & 99&2 \\
    syslog & 85.1 & 99&1.5 \\
    udp-storm & 86.4 & 210.1&2.5 \\
    custom$^*$ & 112.7 & 99&2.75 \\ 
    \hline
    http-tunnel & 109 & 526.1&1.2 \\
    sendmail & 117.3 & 657.2&2 \\ 
    \hline
    
    sql & 111 & 488.8&2 \\
    ffbconfig & 112 &no &1.4 \\
    ps & 110 &no &2 \\
    \hline
    \hline
    
    \textbf{Average} & 103.6 & 311.3&1.8 \\
    \hline
    \end{tabular}
}
\end{table}



\section{Conclusion and Future Work}
Overall, our results suggest that R-ABS provides a significant improvement over the ABS model in terms of quality and classification speed. These advantages are achieved by the learning capabilities and the control of the adaptive conditions for the characterizing agents given by R-ABS. We provide a test of the R-ABS capabilities in the security domain, allowing the comparison with the original ABS model. However, R-ABS is thought of as a generalist tool for anomaly detection and characterization, useful for other domains like complex machine failure management. The exploration of new application domains and the improvement of the different components of R-ABS are the future works for this research.

\bibliographystyle{plainnat}
\bibliography{refs}   


%
%

\end{document}